# BOTDA using OFDM channel estimation

Can Zhao,[1] Ming Tang,[1,4] Liang Wang,[2,5] Hao Wu,[1] Zhiyong Zhao,[1] Yunli Dang,[1] Jiadi Wu,[1] Songnian Fu,[1] Deming Liu,[1] and Perry Ping Shum[3]

[1]*Wuhan National Lab for Optoelectronics (WNLO) & National Engineering Laboratory for Next Generation Internet Access System, School of Optics and Electronic Information, Huazhong University of Science and Technology, Wuhan 430074, China*
[2]*Department of Electronic Engineering, The Chinese University of Hong Kong, Shatin, N.T. Hong Kong*
[3]*School of Electrical and Electronics Engineering, Nanyang Technological University, 637553, Singapore*
[4]*tangming@mail.hust.edu.cn*
[5]*lwang@ee.cuhk.edu.hk*

**Abstract:** A novel Brillouin optical time-domain analysis (BOTDA) system is proposed using intensity-modulated optical orthogonal frequency division multiplexing probe signal and direct detection (IM-DD-OOFDM) without frequency sweep operation. The influence of peak to average power ratio (PAPR) of OFDM probe signal on the recovery of Brillouin gain spectrum (BGS) is analyzed in theory and experiment. The complex BGS is reconstructed by channel estimation algorithm and Brillouin frequency shift (BFS) is located by curve fitting of intensity spectrum. The IM-DD-OOFDM BOTDA is demonstrated experimentally with 25m spatial resolution over 2 km standard single mode fiber.

**Keyword**s: stimulated Brillouin Scattering, Fiber optics sensors

## References and links

1. T. Horiguchi, K. Shimizu, T. Kurashima, M. Tateda, and Y. Koyamada, "Development of a distributed sensing technique using Brillouin scattering," J. Lightwave Technol. **13**(7), 1296–1302 (1995).
2. Y. Peled, A. Motil, L. Yaron, and M. Tur, "Slope-assisted fast distributed sensing in optical fibers with arbitrary Brillouin profile," Opt. Express **19**(21), 19845–19854 (2011).
3. Y. Peled, A. Motil, and M. Tur, "Fast Brillouin optical time domain analysis for dynamic sensing," Opt. Express **20**(8), 8584–8591 (2012).
4. Y. Peled, A. Motil, I. Kressel, and M. Tur, "Monitoring the propagation of mechanical waves using an optical fiber distributed and dynamic strain sensor based on BOTDA," Opt. Express **21**(9), 10697–10705 (2013).
5. A. Voskoboinik, W. Jian, B. Shamee, S. R. Nuccio, L. Zhang, M. Chitgarha, A. E. Willner, and M. Tur, "SBS Based Fiber Optical Sensing Using Frequency-Domain Simultaneous Tone Interrogation," J. Lightwave Technol. **29**(11), 1729–1735 (2011).
6. A. Voskoboinik, O. F. Yilmaz, A. W. Willner, and M. Tur, "Sweep-free distributed Brillouin time-domain analyzer (SF-BOTDA)," Opt. Express **19**(26), B842–B847 (2011).
7. C. Jin, N. Guo, Y. Feng, L. Wang, H. Liang, J. Li, Z. Li, C. Yu, C. Lu, "Scanning-free BOTDA based on ultra-fine digital optical frequency comb," Opt. Express **23**(4), 5277–5284 (2015).
8. J. Fang, P. Xu, and W. Shieh, "Single-shot measurement of stimulated Brillouin spectrum by using OFDM probe and coherent detection," in *Photonics and Fiber Technology 2016 (ACOFT, BGPP, NP)*, OSA Technical Digest (online) (Optical Society of America, 2016), paper AT5C.3.
9. X. Yi, W. Shieh, and Y. Ma, "Phase Noise Effects on High Spectral Efficiency Coherent Optical OFDM Transmission," J. Lightwave Technol. **26**(10), 1309–1316 (2008).
10. W. Shieh, R. S. Tucker, W. Chen, X. Yi, and G. Pendock, "Optical performance monitoring in coherent optical OFDM systems," Opt. Express **15**(2), 350–356(2007).
11. X. Yi, Z. Li, Y. Bao, and K. Qiu, "Characterization of Passive Optical Components by DSP-Based Optical Channel Estimation," IEEE Photon. Technol. Lett. **24**(6),443-445 (2012).
12. B. Guo, T. Gui, Z. Li, Y. Bao, X. Yi, J. Li, X. Feng, and S. Liu, "Characterization of passive optical components with ultra-fast speed and high-resolution based on DD-OFDM," Opt. Express **20**(20), 22079-22086 (2012).
13. L. Tao, Y. Ji, J. Liu, A. P. T. Lau, N. Chi, and C. Lu, "Advanced modulation formats for short reach optical communication systems," IEEE Netw.**27**(6), 6–13 (2013).
14. C. Wei, H. Y. Chen, H. Chu, Y. Chen, C. Song, I. Lu, and J. Chen, "32-dB Loss Budget High-Capacity OFDM Long-Reach PON over 60-km Transmission without Optical Amplifier," in *Optical Fiber Communication Conference*, OSA Technical Digest (online) (Optical Society of America, 2014), paper Th3G.1.

15. C. Tellambura, "Upper bound on peak factor of N-multiple carriers," Electronics Letters. **33**(19), 1608-1609 (1997).
16. Marcelo A. Soto and Luc Thévenaz, "Modeling and evaluating the performance of Brillouin distributed optical fiber sensors," Opt. Express **21**(25), 31347-31366 (2013).
17. L. Thévenaz, S. F. Mafang, and J. Lin, "Effect of pulse depletion in a Brillouin optical time-domain analysis system," Opt. Express **21**(12), 14017-14035 (2013).
18. Z. Feng, M. Tang, S. Fu, l. Deng, Q. Wu, R. Lin, R. Wang, P. Shum, D. Liu, "Performance-enhanced direct detection optical OFDM transmission with CAZAC equalization," IEEE Photon. Technol. Lett. **27**(14),1507-1510 (2015).
19. A. W. Brown, M. D. DeMerchant, X. Bao, T. W. Bremner, "Precision of a Brillouin-scattering-based distributed strain sensor," Proc. SPIE **3670**, 359 (1999).

## 1. Introduction

During the past decades, optical fiber sensing based on Brillouin scattering has been turned into one of the most remarkable fields of research in optical fiber sensing, mainly due to its ability of distributed temperature and strain measurements along tens of km of optical fiber [1]. Classical BOTDA system needs to conduct a large number of averaging process to enhance the signal to noise ratio (SNR) with measuring time of ~minute level, which is unsuitable for dynamic distributed sensing. To improve the acquisition speed, several techniques have been proposed. For instance, slope assisted technique is realized by tuning the frequency of probe optical wave to the position with half of the peak intensity in Brillouin gain spectrum (BGS), in such way, BFS is related to gain variations of probe wave [2-4]. However, the dynamic range is limited by the linear range of Brillouin gain. Another sweep-free BOTDA based on multiple probe and pump pulse pairs is proposed [5,6]. These probe tones are arranged in such a way that each probe tone is located in a different region of the BGS of the corresponding pump pulse. Nevertheless, BFS that moves larger than the frequency spacing between adjacent pump tones will introduce ambiguity, thus severely limiting dynamic range. In references [7,8], the scanning-free BOTDA is realized based on an ultra-fine digital optical frequency comb and coherent optical OFDM (C-OOFDM) probe, respectively, but both schemes suffer from rather high complexity and cost. In fact, the traditional frequency-sweep operation can be seen as a kind of discrete spectrum analysis of probe signal, from which point, channel estimation methods show great potential for BGS recovery. On the other hand, digital signal processing (DSP) has significantly improved optical fiber transmission by advanced algorithm design. The optical orthogonal frequency division multiplexing (OOFDM) is a typical example to take advantage of DSP for optical channel estimation, which has been utilized to calculate the combined transfer function of many optical components [9,10].

In this paper, we propose a novel BOTDA system using optical orthogonal frequency division multiplexing probe signal and direct detection(DD-OOFDM) channel estimation, inspired by [11,12]. The sweep-free BOTDA system is demonstrated experimentally with discrete multi-tone (DMT) OFDM modulated probe and direct detection. The impact of peak to average power ratio (PAPR) of OFDM signal on BGS reconstruction is theoretically analyzed and experimentally verified. At last, distributed stimulated Brillouin spectrum is measured in a 2km single-mode fiber within much less measuring time with a spatial resolution of about 25m.

## 2. Theory of BGS measurement based on channel estimation

In linear system, the input and output signals of an optical component are related by its transfer function. That is to say, the performance of an optical fiber system is determined by its complete transfer function. In BOTDA system, the probe signal before and after interaction with the pump pulse can be regarded as input and output signals, respectively. Thus, the transfer function (complex BGS) could be calculated by optical channel estimation.

In our proposed scheme, channel estimation is conducted in the frequency domain within OFDM signal. In general, the transmitted OFDM signal can be described as follows:

$$s(t) = e^{j2\pi f_0 t} + \alpha \cdot e^{j2\pi f_0 t} \cdot s_B(t) \tag{1}$$

where $s(t)$ is the optical OFDM signal, $f_0$ is the frequency of optical carrier, $\alpha$ is the scaling efficiency indicating the relative amplitude between the optical carrier and OFDM signal. $s_B(t)$ is the baseband OFDM signal given by Eq. (2):

$$s_B(t) = \sum_{k=0}^{N-1} c_k e^{j2\pi f_k t} \tag{2}$$

where $c_k$ represents the mapped symbols on the $k$th subcarrier, $f_k$ and $N$ represent the frequency of the $k$th subcarrier and total number of subcarriers respectively. Let $h(t)$ be the impulse response function of the system, then the received OFDM signal can be approximated using Eq. (3):

$$\begin{aligned} r(t) &= \int_{-\infty}^{\infty} s(t-\tau)h(\tau)d\tau \\ &= e^{j2\pi f_0 t}\int_{-\infty}^{\infty} h(\tau)\cdot e^{-j2\pi f_0 \tau}d\tau + \alpha \cdot e^{j2\pi f_0 t}\cdot \sum_{k=0}^{N-1} c_k e^{j2\pi f_k t}\int_{-\infty}^{\infty} h(\tau)e^{-j2\pi(f_0+f_k)\tau}d\tau \end{aligned} \tag{3}$$

Note that, channel response function $H(f)$ can be written as the Fourier transform of the impulse response function:

$$H(f) = \int_{-\infty}^{\infty} h(\tau)\cdot e^{-j2\pi f \tau}d\tau \tag{4}$$

Thus, Eq. (3) can be simplified as follows:

$$r(t) = H(f_0)e^{j2\pi f_0 t} + \alpha \cdot e^{j2\pi f_0 t}\cdot \sum_{k=0}^{N-1} H(f_0+f_k)c_k e^{j2\pi f_k t} \tag{5}$$

At the receiver, the output current is determined by the square law and can be represented as Eq. (6):

$$\begin{aligned} I(t) \propto |r(t)|^2 &\approx |H(f_0)|^2 + 2\alpha \cdot H(f_0)^* \cdot \mathrm{Re}\left\{\sum_{k=0}^{N-1}\left[\underbrace{\left[|H(f_0+f_k)|e^{j\phi(f_k)}\cdot c_k\right]}_{r_k} e^{j2\pi f_k t}\right]\right\} \\ &+ \alpha^2 \left(\sum_{k=0}^{N-1} H(f_0+f_k)c_k e^{j2\pi f_k t}\right)\left(\sum_{k=0}^{N-1} H(f_0+f_k)^* c_k^* e^{-j2\pi f_k t}\right) \end{aligned} \tag{6}$$

In Eq. (6), the first term is DC component which will be removed during DSP process. The second term contains the received OFDM signal with both the phase and intensity response information. Let $r_k$ be the received mapped data, then the complex channel estimation can be calculated by $H(f_0+f_k) = r_k/c_k$ and the intensity response is the absolute value of $H(f_0+f_k)$. For the third term, $\alpha$ can be set to make $|\alpha|^2$ much smaller, so it can be ignored. By calculating $H(f_0+f_k)$ for each subcarrier, the complex spectrum can be obtained.

In our proposed BOTDA system, the optical OFDM signal serves as the probe signal. If we denote $r_k'$ as the received mapped data after interaction with pump pulse, the complex response becomes:

$$H(f_0+f_k)' = \frac{r_k'}{c_k} = H(f_0+f_k)H_{SBS} \tag{7}$$

where $H_{SBS}$ is the complex Brillouin gain spectrum, described by:

$$H_{SBS} = \exp(G_{SBS} + j\varphi_{SBS}) \approx (1 + G_{SBS})\exp(j\varphi_{SBS})$$
$$= \left(1 + \frac{g_B \Delta v_B^2}{\Delta v_B^2 + 4\Delta v^2}\right) \exp\left(-j \frac{2g_B \Delta v \Delta v_B}{\Delta v_B^2 + 4\Delta v^2}\right) \quad (8)$$

with $G_{SBS}$ and $\varphi_{SBS}$ as the Brillouin gain and phase shift, respectively, under a small gain regime for BOTDA sensors, $g_B$ is the peak gain, $\Delta v_B$ is the Brillouin linewidth, $\Delta v$ is the detuning of the interaction frequency from the center of the Brillouin spectrum. Eq. (8) shows that both BGS and Brillouin phase spectrum (BPS) can be easily obtained through channel estimation.

## 3. Influence of OFDM PAPR on BGS recovery

DD-OOFDM system attracts great attention in short reach optical communication systems because of its high spectral efficiency and robustness against fiber dispersion. However, DD-OOFDM system suffers from the large PAPR that may result in significant distortions to the transmitted signal, leading to power efficiency degradation [13]. Consequently, the optical modulation index of an IM-DD based system is restrained to maintain the linearity, resulting in worse receiver sensitivity and insufficient optical power budget [14].

For the baseband OFDM signal in Eq. (2), the PAPR can be defined as

$$PAPR(dB) = 10\log_{10} \frac{\max\left(|s_B(t)|^2\right)}{E\left\{|s_B(t)|^2\right\}} \quad (9)$$

If the amplitude of all subcarriers is normalized, the average power of OFDM signal is $E\{|s_B(t)|^2\} = N$. Therefore,

$$PAPR(dB) = 10\log_{10} \frac{\max\left(|s_B(t)|^2\right)}{N} \quad (10)$$

The instantaneous power of $s_B(t)$ is

$$P(t) = |s_B(t)|^2 = s_B(t) \cdot s_B^*(t)$$
$$= \sum_{i=0}^{N-1}\sum_{k=0}^{N-1} c_i c_k^* \exp\left[j2\pi f_{(i-k)}t\right]$$
$$= N + 2\mathrm{Re}\left\{\sum_{i=0}^{N-2}\sum_{k=i+1}^{N-1} c_i c_k^* \exp\left[j2\pi f_{(i-k)}t\right]\right\} \quad (11)$$
$$= N + 2\mathrm{Re}\left\{\sum_{m=1}^{N-1}\exp(j2\pi f_m t)\sum_{i=0}^{N-1-m} c_i c_{(i+m)}^*\right\}$$

For any complex z, $\mathrm{Re}(z) \leq |z|$, $\left|\sum z_n\right| \leq \sum |z_n|$. Therefore,

$$P(t) \leq N + 2\sum_{m=1}^{N-1}|\rho(m)| \quad (12)$$

where $\rho(m) = \sum_{i=0}^{N-1-m} c_i c_{(i+m)}^*$, $m = 0,1,...,N-1$ is the aperiodic autocorrelation function. Eq. (12) shows that if the aperiodic autocorrelation modulus of inverse fast Fourier transform (IFFT) operation input sequence $c_k$ is small (i.e. small $|\rho(m)|$ for $m \geq 1$), the peak-power factor of the signal obtained by passing through the multi-carrier combination can also be small [15]. The peak value of the autocorrelation is the average power of input sequence. Then if the number of subcarriers is not changed, this peak value depends on the input sequence. IFFT operation can be viewed as multiplying sinusoidal functions to the input sequence, summing and sampling the results. Thus, the high correlation property of IFFT input causes the sinusoidal functions to be arranged with in-phase form. As a consequence, the sum of these in-phase functions might have a large amplitude. In other words, the PAPR can be reduced by special design of input sequence $c_k$ with smaller autocorrelation modulus.

In practice, high PAPR means nonuniform distribution of OFDM probe signal power in time domain. On one hand, the limited optical power leads to a lower SNR of the OFDM probe. On the other hand, nonuniform power may cause lower Brillouin gain along certain fiber section. Both factors affect BGS recovery. If the Brillouin gain configuration is considered for simplicity, the CW probe signal is amplified by stimulated Brillouin scattering (SBS) while the pulsed pump propagates along the fiber, as shown in Fig.1. The Brillouin gain induced probe power increase at fiber location z, is given by [16]:

$$\Delta P_s(z) = P_s(z) \left[ \exp\left( \frac{g_B(z)}{A_{eff}} P_p(z) \Delta z \right) - 1 \right] \qquad (13)$$

where $g_B(z)$ is the local Brillouin gain coefficient, $A_{eff}$ represents the nonlinear effective area of the fiber and $\Delta z$ is the interaction length corresponding to the spatial resolution and determined by the pump pulse width.

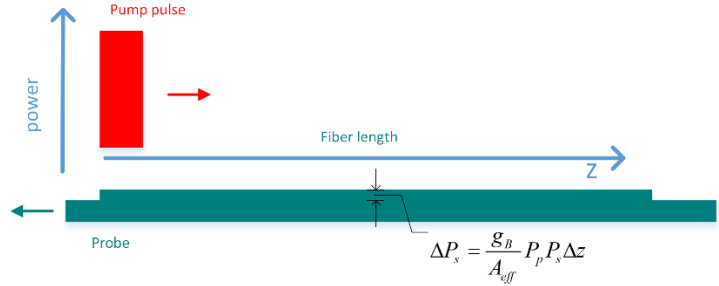

Fig.1. Simplified BOTDA. Pump pulse (with power of $P_p$) interacts with continuous probe (with power of $P_s$) through stimulated Brillouin effect and gives rise to a local power transfer $\Delta P_s$.

The power increase expressed by Eq. (13) actually relates to the local sensor response, where the local gain value $g_B(z)$ is considered for a given frequency difference and is assumed to be constant over the interaction length $\Delta z$. Usually, BOTDA sensors operate in a small gain regime [17], i.e. $\Delta P_s(z)/P_s(z) \ll 0.1$. Therefore, the local sensor response defined by Eq. (13) can be simplified as:

$$\Delta P_s(z) = \frac{g_B(z)}{A_{eff}} P_p(z) P_s(z) \Delta z \qquad (14)$$

In Eq. (14), the local transferred power $\Delta P_s(z)$ from pump is proportional to the probe power $P_s(z)$, which means the SNR is determined by $P_s(z)$. Although, this relation is derived on the basis of single-frequency probe light, it is still applicable to multi-frequency OFDM probe signal. Substitute Eq. (11) into Eq. (14), then the transferred power becomes:

$$\Delta P(z) = \frac{\overline{g_B(z)}}{A_{eff}} P_p(z) P(z) \Delta z \qquad (15)$$

Since $P(z)$ consists of multi-frequency signal and $g_B(z)$ is related to frequency, $\overline{g_B(z)}$ is introduced as constant regardless of frequency difference for simplicity. Different from CW probe, the power of $P(z)$ varies with z, which means the transferred power, $\Delta P(z)$ changes as well. Especially with high PAPR, the Brillouin gain may be quite small at most of the fiber locations, resulting in poor SNR. For BGS measurement in BOTDA, BFS is estimated by curve fitting algorithm. The propagated error on the estimated resonance central frequency $\sigma_v(z)$ can be expressed as [16]:

$$\sigma_v(z) = \sigma(z)\sqrt{\frac{3\cdot\delta\cdot\Delta v_B}{8\sqrt{2}(1-\eta)^{3/2}}} = \frac{1}{SNR(z)}\sqrt{\frac{3\cdot\delta\cdot\Delta v_B}{8\sqrt{2}(1-\eta)^{3/2}}} \quad (16)$$

where $\sigma(z)$ is noise amplitude, directly represented by the inverse of the local SNR at peak gain frequency, $\Delta v_B$ is the estimated full-width at half maximum (FWHM) of the resonance, $\delta$ denotes the frequency spacing, and $\eta$ is the fraction of the peak level over which a quadratic least-square fitting is carried out. Eq. (16) illuminates that the performance of a certain BOTDA system is determined by SNR, which is proportional to the probe power. Since higher PAPR causes severe nonuniform distribution of probe power, which further affects the SNR distribution along fiber, $SNR(z)$, greater error will be introduced in the central frequency estimation. Thus, it can be inferred that the PAPR can impose great influence to OFDM based system, which will be demonstrated in experiments bellow.

## 4. Experimental setup

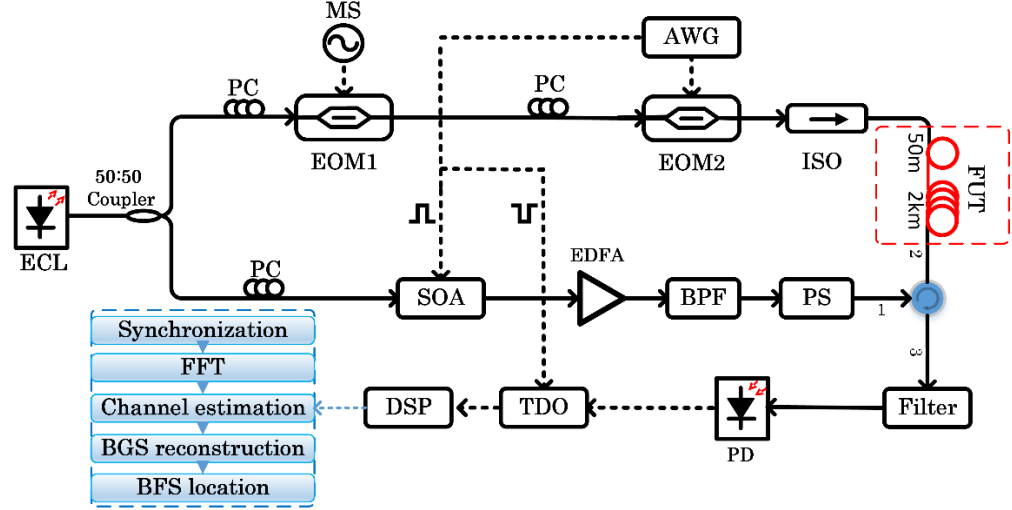

Fig.2. Experimental setup: ECL: external cavity laser, PC: polarization controller, EOM: electro-optic modulator, MS: microwave source, AWG: arbitrary waveform generator, ISO: optical isolator, SOA: semiconductor optical amplifier, EDFA: Erbium doped fiber amplifier, BPF: optical band-pass filter, PS: polarization controller, FUT: fiber under test, PD: photo detector, TDO: time-domain oscilloscope, DSP: digital signal processing.

The experimental setup of IM-DD-OOFDM based BOTDA scheme is shown in Fig.2. The external cavity laser (ECL) working at 1550nm with less than 100 kHz linewidth is divided by a 3dB coupler into two parts as probe and pump respectively. At the probe side, the light is first modulated by an electro-optic intensity modulator (EOM1) driven by a microwave source (MS) with 10.65GHz RF signal to generate dual-sideband optical signal before launched into another intensity modulator (EOM2). Then the OFDM baseband signal is generated from an arbitrary waveform generator (AWG) at the sampling rate of 500MSa/s and added onto dual-sideband probe light by EOM2. Every OFDM subcarrier carries binary phase shift keying (BPSK) mapped Pseudo Random Binary Sequence (PRBS). The OFDM signal has a total number of 128 subcarriers but 63 effective subcarriers, as for real-valued OFDM generation, Hermitian symmetry must be satisfied and the first subcarrier is abandoned to eliminate noise near DC component [18]. Thus, the frequency spacing of adjacent subcarrier is about 4MHz, corresponding to a time duration of 250ns.

The high extinction ratio pump pulse is generated through a semiconductor optical amplifier (SOA), which is driven by a positive pulse with the width of 200ns and period of 30us from another channel of AWG. An inverse electrical signal from the same channel is used as the trigger of the real-time oscilloscope. After amplified by an Erbium doped fiber amplifier (EDFA), the pump pulse with high peak power is launched into fiber through an optical fiber circulator. An optical band pass filter (BPF) is used to filter out amplifier spontaneous emission noise (ASE) of EDFA. A polarization switch (PS) is placed in front of the circulator to alleviate polarization dependent fluctuations of Brillouin gain. At the receiver, the higher frequency sideband of the probe signal is first filtered out by a narrow band filter for gain-mechanism based BOTDA, then a photodetector (PD) with a bandwidth of 400MHz is utilized for direct detection of the probe signal. To verify the performance of our proposed scheme, a spool of 2km standard single mode fiber (SSMF) together with 50m SSMF heated is used as fiber under test (FUT).

## 5. Experimental results

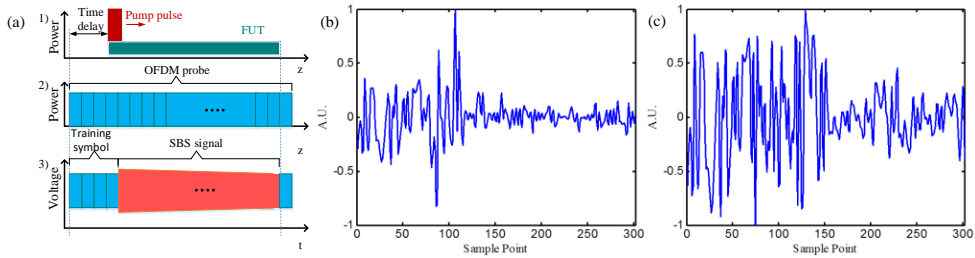

Fig.3. (a) Time-domain signal: 1) Pump pulse, 2) OFDM probe, 3) Received signal with training symbol. (b) Normalized baseband OFDM signal with PAPR of 15dB. (c) Normalized baseband OFDM signal with PAPR of 9dB.

Fig.3 (a) depicts the schematic illustration of our scheme in the time-domain. The first four symbols are designed as training symbols for synchronization and the remaining symbols are identical for simplicity. Since the length of OFDM signal is the same as the pulse period, the last few symbols are kept unaffected and can be used for channel estimation. The received probe OFDM signals are first averaged 512 times to improve the SNR. After that, the received data are divided into many segments, each of which contains one symbol, corresponding to a spatial resolution of 25m. Then fast Fourier Transform (FFT) is applied to every segment to obtain complex amplitude of each subcarrier. The BGS along the FUT can be obtained after processing all the segments. As aforementioned, both the BGS and BPS can be calculated through channel estimation, though we choose BGS for sensing.

To validate the influence of high PAPR of OFDM signal on BGS reconstruction, two different OFDM signals with PAPR of 15dB and 9dB are used for probe wave generation, as shown in Fig.3 (b) and (c), respectively. As analyzed in Section 3, the power distribution with time is quite different for OFDM signal with different PAPR. Higher PAPR leads to fewer power peaks and gives rise to small values in the majority of sample points, resulting in smaller power gain through SBS.

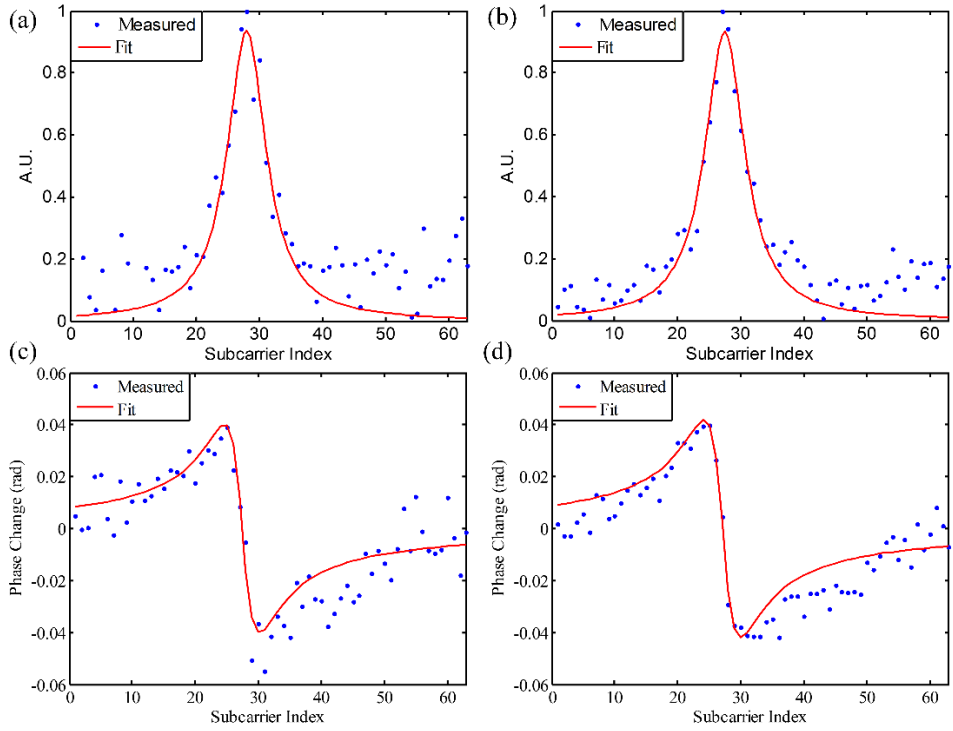

Fig.4. (a) Brillouin intensity gain spectrum for PAPR of 15dB. (b) Brillouin intensity gain spectrum for PAPR of 9dB. (c) Brillouin phase spectrum for PAPR of 15dB. (d) Brillouin phase spectrum for PAPR of 9dB.

The gain spectrum reconstructed in both cases are represented as blue dots in Fig.4 (a) and (b), respectively. As can be seen, the spectrum obtained in higher PAPR case fluctuates more seriously, especially for subcarriers with lower Brillouin gain. The red solid lines in Fig.4 (a) and (b) indicate the respective Lorentzian curve fitting results, which are obtained using data above the level of 0.2 of peak gain. Thanks to higher SNR, the measured spectrum in lower PAPR case fits better to the Lorentzian curve. Similarly, the phase spectrum of lower PAPR performs better, as denoted in Fig.4. (d) compared to that in Fig.4. (c). Since the SNR of peak gain frequency of probe cannot be precisely estimated by calculating error vector magnitude (EVM) as in communication systems, we adopt the method in reference [19] to evaluate the performance of the system. Accordingly, the SNR of the probe is calculated to be 31.3dB and 28.5dB for PAPR of 9dB and 15dB, respectively. And the corresponding estimation error of central frequency for BGS curve fitting is about 0.3MHz and 0.4MHz, respectively. Thus, the performance of the system degrades at high PAPR of OFDM signal. It should be noted that the PAPR is related to the number of subcarriers and the measurement range is in proportion to the bandwidth (i.e. the number of subcarriers for fixed frequency spacing). That is to say, there will be a tradeoff between reducing PAPR and increasing measurement range. Therefore, the problem of high PAPR of OFDM signal should be considered carefully for better performance system design.

To verify the performance of our proposed system, the 50m SSMF spliced at the end of 2km SSMF is kept in a water bath pot heated from 30 ºC to 80 ºC with the step of 10 ºC. At room temperature, the BFSs of the two fiber sections are 10.775GHz and 10.752GHz, respectively. The OFDM signal with PAPR of about 9dB is used for probe in the experiment.

Fig.5 (a) and Fig.5 (b) show the reconstructed BGS and measured BFS of the fiber heated in the water pot, respectively. As depicted in Fig.5 (a), the BGS distribution along FUT consists of two sections, corresponding to the long fiber spool and short spliced fiber tail. Because of the 25m spatial resolution which is dependent on the length of OFDM symbol, it is clear to observe the heated segment as shown by the inset in Fig.5 (a). As the symbol length is inversely proportional to the frequency space of adjacent subcarriers, one can simply increase the frequency space for higher spatial resolution, within an acceptable BFS estimation error ($\sigma_v(z)$).

The measured BFS of heated fiber section as a function of temperature is plotted as blue dots in Fig.5 (b), which is in good agreement with the linear fitting red solid line. The temperature coefficient for the fiber segment is calculated to be 1.03 ºC/MHz.

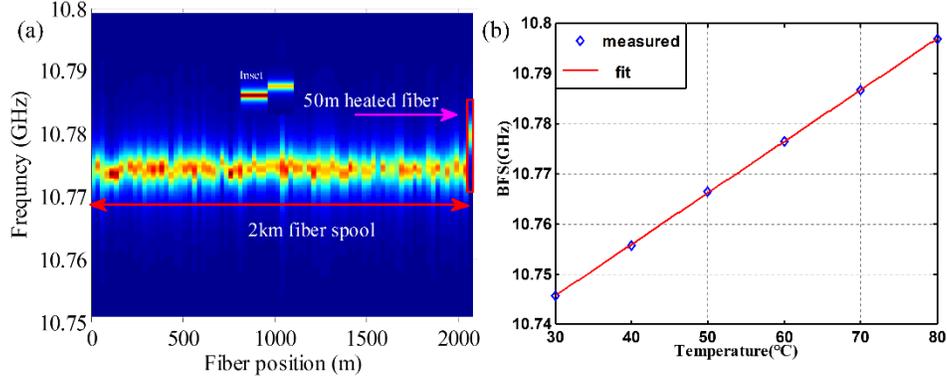

Fig.5. (a) Reconstructed spectrogram along the fiber. (b) Measured and fit BFS as a function of temperature.

In our experiment, the time consumption for each acquisition is about 30*μs*, and 512 times averaging is needed in a single measurement to improve the SNR together with two measurements for each polarization state of the pump pulse, resulting in a total measurement time of about 30*ms*. Since the performance of the system is largely dependent on the SNR of the obtained probe signal and the high PAPR of OFDM probe is one of the main factors that affects the distributed SNR ($SNR(z)$), more work should be done for optimal OFDM signal design.

## 6. Conclusions

We have proposed and experimentally demonstrated a novel sweep-free BOTDA system based on IM-DD-OOFDM channel estimation, which is cost effective with improved measurement efficiency. Through channel estimation algorithm, the complex BGS is obtained and the BFS is located by BGS curve fitting with the uncertainty less than 1MHz. Moreover, the influence of PAPR on system performance is theoretically analyzed and experimentally verified. It implies that OFDM probe should be deliberately designed for better performance Distributed temperature measurement over 2km SSMF is conducted and the total measurement time is within tens of milliseconds, which means the proposed scheme has the obvious advantage in time saving over conventional BOTDA system.

### Funding

This work is supported by the National Natural Science Foundation of China (NSFC) under Grant No. 61331010, the 863 High Technology Plan of China (2013AA013402) and the Program for New Century Excellent Talents in University (NCET-13-0235).